\documentclass[12pt]{iopart}

\usepackage{graphicx}

\begin{document}

\title{General solution for scalar perturbations in bouncing cosmologies}

\author{V. Bozza}

\address{Dipartimento di Fisica ``E. R. Caianiello",
Universit\`a di Salerno, I-84081 Baronissi, Italy. \\
 Istituto Nazionale di Fisica Nucleare, Sezione di Napoli,
 Naples, Italy.}

\begin{abstract}
Bouncing cosmologies, suggested by String/M-theory, may provide an
alternative to standard inflation to account for the origin of
inhomogeneities in our universe. The fundamental question regards
the correct way to evolve the scalar perturbations through the
bounce. In this work, we determine the evolution of perturbations
and the final spectrum for an arbitrary (spatially flat) bouncing
cosmology, with the only assumption that the bounce is governed by
a single physical scale. In particular, we find that the spectrum
of the pre-bounce growing mode of the Bardeen potential (which is
scale-invariant in some limit, and thus compatible with
observations) survives unaltered in the post-bounce only if the
comoving pressure perturbation is directly proportional to the
Bardeen potential rather than its Laplacian, as for any known form
of ordinary matter. If some new physics acting at the bounce
justifies such relation, then bouncing cosmologies are entitled to
become a real viable alternative for the generation of the
observed inhomogeneities. Our treatment also includes some class
of models with extra-dimensions, whereas we show that bounces
induced by positive spatial curvature are structurally different
from all bounces in spatially flat universes, requiring a distinct
analysis.
\end{abstract}


\maketitle

\section{Introduction}

The occurrence of singularities in classical General Relativity
(GR) may be the symptom of the existence of a more general
physical frame, where GR is embedded as a particular low curvature
limit. In such a frame, that we will generically call ``new
physics'', one may expect that singularities are cured by the
existence of a natural physical cutoff, i.e. a scale where the new
physics comes into play, replacing GR. Presently, the only
consistent example of such a general theory is String theory, or
its 11-dimensional generalization M-theory, where this cutoff is
naturally provided by the string length $l_s$.

Following this reasoning line, in a cosmological context, it is
natural to expect that the new physics would prevent the Big Bang
singularity predicted by GR, allowing a continuation of the
cosmological solutions before the Big Bang. This idea, firstly
proposed by Veneziano and Gasperini \cite{PBB}, has found several
alternative implementations and recurrently emerges in independent
attempts to build consistent cosmological models from
String-M/theories \cite{OtherBounce,Ekp,FinBra,Biswas}. All these
models, when formulated in the Einstein frame, share the same
general behavior: the universe starts from an asymptotically flat
state and undergoes a contraction that pumps the space-time
curvature up to the new physics scale. At this point, the new
physics masters the transition to an expanding phase and the
standard Friedman-Robertson-Walker (FRW) universe begins. In the
present literature, such models are generically called {\it
bouncing cosmologies}.

The interest in bouncing cosmological models comes from the fact
that an accelerated contraction seems to solve the flatness and
horizon problems as efficiently as a standard inflationary phase.
Then the question opens whether a bouncing cosmology may replace
standard inflation once and for all. In order to achieve this
fascinating objective, a bouncing cosmology should provide its own
mechanism of generation of the inhomogeneities of the universe, in
such a way that all present constraints are satisfied at the same
accuracy level of standard inflation. The most stringent
observational constraint comes from CMB anisotropies, which
require the initial power spectrum of the Bardeen potential
(encoding the information on all scalar cosmological
perturbations) to be (nearly) scale-invariant. Is this obtainable
or not in a bouncing cosmology? The first studies in the Pre-Big
Bang scenario gave a negative answer \cite{PBBPert}, indicating
that the inhomogeneities must have an independent origin. For
example, they can be generated in an axion-dominated era through
the curvaton mechanism \cite{BGGV}.

On the other hand, the authors of the Ekpyrotic/cyclic scenario
\cite{Ekp} claimed that a scale-invariant spectrum can be
generated in a slow contraction era and transmitted across the
bounce. This has risen a considerable debate \cite{Others,DurVer}
about the correct way to describe the evolution of perturbations
through a bounce, but, after several years of investigation, it
seems quite difficult to draw a general conclusion, valid for all
models of bouncing cosmologies.

The evolution of perturbations in a bouncing cosmology can be
summarized very quickly as follows: in the pre-bounce phase, the
Bardeen potential has a growing mode, endowed with a red spectrum
(becoming scale-invariant in the limit of very slow contraction
\cite{Ekp}) and a decaying mode with a blue spectrum (becoming
scale-invariant in the limit of dust-dominated contraction
\cite{FinBra}). After the bounce, we have two alternatives: either
the spectrum of the growing mode is inherited by a constant mode
in the post-bounce or the growing mode simply decays in the
post-bounce and some other constant mode with a more blue spectrum
dominates at horizon re-entry.

If the first possibility occurs, then the Ekpyrotic/cyclic models
would become a complete alternative to standard inflation.
Otherwise, a curvaton mechanism or some other one must be
implemented to make such models compatible with observations. All
studies of specific (spatially flat) toy models, where
perturbations are explicitly calculable, show that the pre-bounce
growing mode decays in the post-bounce and the Bardeen potential
is finally dominated by a blue spectrum, incompatible with
observations \cite{Cartier,GGV,AllWan,BozVen} (the incorrect
conclusions reached in Refs. \cite{PetPin,Fin} are revised in Ref.
\cite{BozVen} and reconciled with all other studies; also the
model presented in Ref. \cite{BatGes} is a particular case of the
general model presented in Ref. \cite{BozVen}, with two perfect
fluids with $w_a=1/3$ and $w_b=5/3$). However, the investigation
of specific models does not allow to infer any general
conclusions, since the possibility that some still unknown new
physics may intervene at the bounce, producing different effects,
remains open. Indeed, bounces induced by positive spatial
curvature show a different behavior, with a generic mixing between
growing and decaying modes \cite{K>0,Deruelle}.

The description of the evolution of scalar perturbations through a
bounce has acquired an independent interest that goes beyond the
Ekpyrotic/cyclic models, thanks to the mathematical complexity of
the problem, which poses several questions and issues of general
theoretical interest.

In this work we present a completely analytical study of scalar
perturbations in spatially flat bouncing cosmologies, making the
only assumption that the bounce is governed by a single physical
scale. In this way, we are able to establish a univocal relation
between the final spectrum of perturbations and the form of the
total energy-momentum tensor. We thus introduce a sort of golden
rule for scalar perturbations in bouncing cosmologies.

In Sect. 2 we define the generic bouncing cosmology, stating the
hypotheses that we need to deduce our general result. In Sect. 3
we elaborate our hypotheses, describing the background cosmology.
In Sect. 4 we write the linear perturbation equations. In Sect. 5
we derive the general form of the effective energy-momentum
tensor, including the high energy corrections driving the bounce.
In Sect. 6 we compute the evolution of perturbations in the
pre-bounce, in the bounce and in the post-bounce. In Sect. 7 we
illustrate the general results of Sect. 6 by some numerical
examples. In Sect. 8 we consider possible extensions to models
with extra-dimensions and analyze the differences between bounces
in spatially flat universes and bounces in spatially closed
universes. Sect. 9 contains the conclusions.

\section{The generic bouncing cosmology}

Let us start by enumerating our hypotheses:

\begin{itemize}
\item[(i)] We assume that it makes sense to define a four-dimensional metric
tensor, both in the GR regime and in the new physics regime. Then
it is possible to write the four-dimensional Einstein equations
with an effective energy-momentum tensor on the right hand side,
encoding all the corrections dictated by the new physics.

\item[(ii)] The universe is homogeneous, isotropic and spatially flat,
thus being described by the FRW metric.

\item[(iii)] There is a unique physical scale $\eta_{B}$ governing
the bounce. We will set our units in such a way that $c=1$, $8\pi
G=1$, $\eta_B=1$.

\item[(iv)] In the GR regime far from the bounce ($|\eta|\gg 1$),
the energy momentum tensor is characterized by a constant equation
of state and a constant speed of sound, with $p/\rho>-1/3$.
\end{itemize}

In Sect. 8, we will comment on models with extra-dimensions which
may fit into hypothesis (i) and discuss the differences with
positive curvature bounces.

Now let us work out these hypotheses. The metric, including scalar
perturbations, is
\begin{equation}
ds^2=a^2(\eta)\left\{ (1+2\phi) d\eta^2-2B_{,i} d\eta dx^i
-\left[(1-2\psi)\delta_{ij}+2E_{,ij}\right]dx^i dx^j \right\},
\end{equation}
where we have used the conformal time $\eta$. The scale factor is
$a(\eta)$, while $B$, $E$, $\psi$ and $\phi$ are scalar
perturbations.

The total energy--momentum tensor reads
\begin{equation}
{{T}^\mu}_\nu= \left(%
\begin{array}{cc}
  \rho+\delta \rho & -(\rho+p) \mathcal{V}_{,i} \\
  (\rho+p) \mathcal{V}_{,i} & -(p+\delta p) \delta_{ij} -\xi_{,ij}\;  \\
\end{array}%
\right), \label{TmunuP}
\end{equation}
where $\delta \rho$ is the energy density perturbation, $\delta p$
is the pressure perturbation, $\mathcal{V}$ is the velocity
potential and $\xi$ is the anisotropic stress. Note that these
expressions do not require GR, but are just dictated by the
symmetry of the problem (hypothesis (ii)). Thus, also in the
bounce regime, the new physics corrections encoded in
${T^\mu}_\nu$ will respect this symmetry and contribute to its
components, both at the background and at the perturbative level.
Since this tensor stays on the right hand side of the Einstein
equations, it must be compatible with the Bianchi identities, thus
it must be covariantly conserved: ${T^\mu}_{\nu;\mu}=0$.

The independent background Einstein equations and the
energy-momentum conservation give the Friedman equations
\begin{eqnarray}
&& 3\mathcal{H}^2=a^2 \rho \\ && \mathcal{H}^2+2\mathcal{H}'=-a^2
p
\\ && \rho'+3\mathcal{H}(\rho+p)=0, \label{Eqrho'} \\ \nonumber
\end{eqnarray}
where the prime denotes derivative w.r.t. $\eta$ and
$\mathcal{H}=a'/a$ is the conformal Hubble rate.

In a spatially flat universe the Null Energy Condition (NEC),
which amounts to require $\rho+p>0$, must be necessarily violated
during the bounce. Of course, this violation cannot be achieved by
ordinary matter, but requires the intervention of the new physics.
One may object that this implies the presence of ghosts at the
quantum level. While the majority of explicit bouncing cosmologies
in the literature indeed contain ghosts, there are at least two
counterexamples. The first is the model by Gasperini, Giovannini
and Veneziano \cite{GGV}, where the NEC violation is performed by
a potential for the scalar field involving integration on the
spatial volume of the universe. The second is the model by Biswas,
Mazumdar and Siegel \cite{Biswas}, where the generalized
gravitational action is an analytic function of the D'Alembert
operator acting on the curvature scalar without poles. In both
cases there are no ghosts in the theory. In any case, the possible
existence of ghosts for a finite time interval may not be a
problem at all \cite{Starobinsky}.

\section{Background}

Now let us determine the background evolution far from the bounce,
using hypothesis (iv). In the pre-bounce we have $p=w_- \rho$ and
in the post-bounce $p=w_+\rho$, with constant $w_-$ and $w_+$. The
restriction $w_+ >-1/3$ excludes the possibility of a post-bounce
inflation, which would spoil the results of the bounce. On the
other hand, a pre-bounce deflation, excluded by the restriction
$w_- >-1/3$, would make the setting of initial conditions in the
asymptotic past problematic.

We set $\eta=0$ at the bounce, defined as the time where
$a'(\eta)=0$. The bounce duration scale must be of the same order
of the curvature scale, according to hypothesis (iii). Otherwise,
there would be two independent scales. So, the bounce will last
from $\eta_- \simeq -1$ to $\eta_+ \simeq 1$.

Then, the pre-bounce background functions have the power law
behavior
\begin{eqnarray}
&& a(\eta)\simeq |\eta|^{q_-} \label{PowLawa}\\ &&
\mathcal{H}\simeq\frac{q_-}{\eta} \\ && \rho_a\simeq
\frac{3q_-^2}{|\eta|^{2+2q_-}} \\ && q_-=\frac{2}{1+3w_-},
\label{PowLawq}
\end{eqnarray}
where we have chosen all constant of integration in such a way
that the space-time curvature at the onset of the bounce $\eta
\sim -1$ is of order 1.

In the same way, we can write the post-bounce behavior of the
background functions, which will be analogous to Eqs.
(\ref{PowLawa})-(\ref{PowLawq}) with all ``$-$'' in the subscripts
replaced by ``$+$''.

\section{Perturbations equations}

We will write the perturbations equations in terms of the
following gauge invariant variables

\begin{eqnarray}
&& \Psi=\psi+\mathcal{H}(E'-B) \\
&& \zeta=\psi+\mathcal{H V} \\
&& \delta \rho_\mathcal{V}=\delta \rho- \rho' \mathcal{V} \\
&& \delta p_\mathcal{V}=\delta p- p' \mathcal{V},
\end{eqnarray}
plus the anisotropic stress $\xi$. $\Psi$ is the Bardeen
potential, $\zeta$, $\delta \rho_\mathcal{V}$ and $\delta
p_\mathcal{V}$ are respectively the curvature, energy density and
pressure perturbations on comoving hypersurfaces. In the following
we will drop the subscript ``$\mathcal{V}$'' from all variables.
Then, from now on, $\delta \rho$ and $\delta p$ must be intended
as comoving energy density and pressure perturbations.

Using the component $(0i)$ of the Einstein equations to eliminate
$\phi$, the $(00)$, $(ij)$ and $(ii)$ components become
\begin{eqnarray}
&& \delta \rho=\frac{2}{a^2}\nabla^2 \Psi \label{Eqdrho}\\
&& \delta p
-\nabla^2\xi=-\frac{2(\mathcal{H}^2-\mathcal{H}')}{a^2\mathcal{H}}
\zeta' \label{Eqzeta}\\
&&
a^2\mathcal{H}\xi=-\frac{\mathcal{H}^2-\mathcal{H}'}{\mathcal{H}}
\zeta +\frac{2\mathcal{H}^2-\mathcal{H}'}{\mathcal{H}}\Psi +\Psi'.
\label{EqPsi}
\end{eqnarray}

Some of the coefficients of these equations become singular at the
bounce ($\mathcal{H}=0$) or at the NEC violation time
($\mathcal{H}^2-\mathcal{H}'=0$). In order to check the regularity
of our variables, we may resort to the regular gauge argument used
in Ref. \cite{BozVen}. If a gauge exists where all components of
the metric and energy-momentum tensor stay finite, then we can
explicitly check the regularity of our gauge invariant variables,
expressing them in this gauge. As shown in \cite{BozVen}, $\Psi$
is regular while $\zeta$ is not. However, we may construct
$\tilde{\zeta}=(\mathcal{H}^2-\mathcal{H}')\zeta$, where the extra
factor compensates the possible divergence of $\mathcal{V}$ at the
NEC violation, allowed by the fact that it is the combination
$(\rho+p)\mathcal{V}$ that appears in the energy-momentum tensor
and is guaranteed to be regular by the regular gauge hypothesis.
$\delta \rho_\mathcal{V}$ has no divergences (since $\rho'=0$ at
the NEC violation time, thanks to the continuity equation), while
for $\delta p_\mathcal{V}$ we should use the regularized variable
$\delta\tilde{p}_\mathcal{V}=(\mathcal{H}^2-\mathcal{H}')\delta
p_\mathcal{V}$. Dealing with regularized variables is absolutely
necessary in all numerical applications, because no integrator can
step through a divergence. However, for our analytical arguments,
it is completely equivalent to discuss the equations
(\ref{Eqdrho})-(\ref{EqPsi}) or their regularized counterparts. We
choose the unregularized variables only because they have an
immediate physical interpretation, but we would get exactly the
same results with the regularized variables.

Going back to our equations, Eq. (\ref{Eqdrho}) represents the
analogous of the Poisson equation, relating the Laplacian of the
gravitational potential to the perturbations in the energy density
(this is not a dynamical equation). Eq. (\ref{Eqzeta}) is a first
order equation in $\zeta$, which depends on the sources $\delta p$
and $\xi$. The anisotropic stress also enters as a source in the
first order equation for $\Psi$ (\ref{EqPsi}). It is easy to
verify that the perturbations of the continuity equations give no
new independent equations.

All the dynamics of scalar perturbations is contained in these
three equations. In order to solve them, we need to specify the
sources $\delta p$ and $\xi$ for the energy-momentum content of
our universe. This is easy for the pre-bounce and post-bounce
phases. In fact, if they are dominated by a perfect fluid or a
scalar field with some potential, the sources are
\begin{equation}
\delta p= c_s^2 \delta \rho; ~~~~~ \xi=0,
\end{equation}
with $s$ being ``$-$'' or ``$+$'' in the pre-bounce and
post-bounce respectively. If the squared speed of sound $c_s^2$ is
equal to $w_s$, we have a perfect fluid, otherwise we have a fluid
with some intrinsic isocurvature mode \cite{BozVen}. In
particular, a scalar field with a potential has always $c_s^2=1$,
whatever the equation of state induced by the presence of the
potential. In any case, with these sources, Eqs.
(\ref{Eqdrho})-(\ref{EqPsi}) can be combined into a second order
equation for $\zeta$ or $\Psi$, suitable for vacuum normalization.
Expanding the variables in Fourier modes of wavenumber $k$ as
usual, in the asymptotic past we have
\begin{equation}
\zeta=C
\frac{\mathcal{H}}{a\sqrt{\mathcal{H}^2-\mathcal{H}'}}\sqrt{|\eta|}
H^{(1)}_{\nu}(c_- k |\eta|),
\end{equation}
where $C$ is a normalization constant of order 1, $H^{(1)}_{\nu}$
is the Hankel function of the first kind and $\nu=1/2-q_-$.

We are interested in modes that are well outside the horizon at
the bounce, thus having $k \ll 1$. For such modes, in the range
$-1/k<\eta<-1$, we can use the expansion of the Hankel functions
for small arguments and Eq. (\ref{Eqzeta}) to determine $\Psi$.
Discarding all numerical factors of order one, we have two modes
for each variable
\begin{eqnarray}
&& \zeta \sim k^\nu |\eta|^{1-2q_-} + k^{-\nu}  \label{zetapre}\\
&& \Psi \sim k^{\nu-2} |\eta|^{-1-2q_-} + k^{-\nu}. \label{Psipre}
\end{eqnarray}
$\Psi$ is characterized by a growing mode with a red spectrum
(becoming scale-invariant in the limit $q_-\rightarrow 0$) and a
constant mode with a blue spectrum. $\zeta$ is characterized by a
mode that is growing for $q_->1/2$ and decaying otherwise. This
mode has a spectrum that is blue for $q_-<2$ and red otherwise,
being scale-invariant in the limit of dust-dominated pre-bounce
\cite{FinBra}. The second mode is a blue constant mode.

\section{General form of the energy-momentum tensor of
perturbations}

In order to discuss the evolution of perturbations through the
bounce, we need to specify the form of the sources. In the
ignorance of the new physics governing the bounce, we can still
select all admissible forms of the pressure perturbation and the
anisotropic stress and make a complete discussion of what can be
expected at the end of the bounce.

Firstly, let us deal with the pressure perturbation, leaving aside
the anisotropic stress for the moment.

The pressure perturbation can be a generic linear combination of
$\delta \rho$, $\Psi$, $\zeta$ and a certain number $n$ of
additional internal variables $\delta \varphi_i$, with time
depending coefficients. Of course, the pressure perturbation may
also depend on the time and space derivatives of these variables.
Now, through Eq. (\ref{Eqdrho}), any dependence on $\delta \rho$
can be read as an additional dependence on $\Psi$ and its
derivatives.

For each internal variable it is necessary to specify an
independent linear differential equation in order to close the
system. We are not interested in any new spectral dependence whose
origin is different from vacuum normalization of the original
scalar degree of freedom. Indeed, if such spectral dependences are
carried by these internal variables, they may play an important
role, giving rise to additional isocurvature modes in the
post-bounce, as in the model discussed in Ref. \cite{BozVen}. They
depend on the specific model and must be carefully considered in
any complete treatment. However, our question is more general and
only involves the fate of the original spectrum of $\Psi$ and
$\zeta$. The internal variables may acquire this spectrum only if
$\zeta$, $\Psi$ and their derivatives explicitly appear in their
equations of motion. But then their feedback on $\delta p$ would
be an effective renormalization of the coefficients of $\zeta$,
$\Psi$ and their derivatives in the expression of $\delta p$.

$\delta p$ is then just a linear combination of $\Psi$, $\zeta$
and their derivatives with time-dependent coefficients. However,
$\delta p$ is scalar under spatial rotations and thus it must
contain only scalars. The time derivatives of scalars are still
scalars under spatial rotation, but the spatial derivatives must
be combined with themselves or other vectors of the theory in
order to produce scalars. More specifically, the spatial
derivatives can only appear as $\nabla^2$ or $A^i\partial_i$,
where $A^i$ is a generic 3-vector of the theory. However, since
the background is homogeneous and isotropic, all 3-vectors in the
background must be zero. So, spatial derivatives can only appear
in a Laplacian or a function of the Laplacian.

Summing up, the most general expression for $\delta p$ is
\begin{equation}
\delta p=\frac{2}{a^2}\sum\limits_{n=0}^\infty
F_n(\eta,\partial_\eta)k^{2n}\zeta +
\frac{2}{a^2}\sum\limits_{n=0}^\infty
G_n(\eta,\partial_\eta)k^{2n}\Psi, \label{Gendp}
\end{equation}
where we have supposed that the dependence on the Laplacians is
analytic and allows a regular power expansion and we have
introduced the factor $2/a^2$ for later convenience. Using the
Fourier mode expansion, the Laplacians have been written as
$-k^2$. With the same arguments, a similar expression should also
hold for the anisotropic stress
\begin{equation}
\xi=\frac{2}{a^2}\sum\limits_{n=0}^\infty
J_n(\eta,\partial_\eta)k^{2n}\zeta +
\frac{2}{a^2}\sum\limits_{n=0}^\infty
K_n(\eta,\partial_\eta)k^{2n}\Psi. \label{Genxi}
\end{equation}
Of course, any possible crossed dependence between $\delta p$ and $\xi$
can be easily solved at the linear level.

Then the physics of the model is completely encoded in the
time-dependent operators $F_n(\eta,\partial_\eta)$,
$G_n(\eta,\partial_\eta)$, $J_n(\eta,\partial_\eta)$,
$K_n(\eta,\partial_\eta)$. Let us make some examples.

For a single perfect fluid or a scalar field, as mentioned before,
the only non-vanishing operator is $G_1=-c_s^2$. In the bounce
induced by two perfect fluids or scalar fields, studied in Ref.
\cite{BozVen} and containing the models by Peter \& Pinto-Neto
\cite{PetPin}, Finelli \cite{Fin}, and Allen \& Wands
\cite{AllWan}, two internal variables are present. Taking into
account their dependence on $\zeta$ and $\Psi$, the final
expression for the effective energy-momentum tensor only contains
two non-vanishing coefficients: $F_0$, and $G_1$, i.e. the
pressure perturbation only depends on $\zeta$ and $k^2\Psi$, while
the anisotropic stress is zero. The model by Gasperini, Giovannini
and Veneziano \cite{GGV} is simpler, since it contains no internal
variables and the only non-vanishing operator is $G_1=
4(\mathcal{H}'-\mathcal{H}^2)/{\varphi'}^2$, where $\varphi$ is
the dilaton field in their model. Cartier's model \cite{Cartier}
has a very involved effective energy-momentum tensor with a
non-vanishing anisotropic stress. Using the expressions in the
appendix of his work, we find that the covariant conservation is
fulfilled only in the trivial case $\mathcal{F}=1/2$, $\alpha'=0$,
indicating that some misprint may be present in the expressions
published therein. So, we renounced to identify our operators
$F_n$, $G_n$, $J_n$ and $K_n$ in this otherwise interesting model.

\section{Evolution of perturbations}

In this section we shall study the evolution of scalar
perturbations through all cosmological eras, making no assumptions
on the specific form of the energy-momentum tensor, thus retaining
all terms in the general expressions (\ref{Gendp}), (\ref{Genxi}).

In Ref. \cite{BozVen} an easy way to find the evolution of
perturbations outside the horizon was proposed. It amounts to
write the first order differential equations in their integral
form and write the solution as a recursive series. The integral
form of Eqs. (\ref{Eqzeta})-(\ref{EqPsi}) is

\begin{eqnarray}
&  \Psi &  =
 \frac{\mathcal{H}}{a^2}
\left[ \frac{c_1(k)}{k^2}+ \int \frac{a^2(\mathcal{H}^2-\mathcal{H}')}{\mathcal{H}^2}
\zeta d\eta +\int a^2\mathcal{H}\xi d\eta \right], \label{Psiint}
\\
&  \zeta &  = \left[ c_2(k) -\int
\frac{a^2\mathcal{H}}{2(\mathcal{H}^2-\mathcal{H}')} \delta p
d\eta -\int \frac{a^2\mathcal{H}}{2(\mathcal{H}^2-\mathcal{H}')}
k^2\xi d\eta
 \right], \label{zetaint}
\end{eqnarray}
where $c_1(k)$ and $c_2(k)$ are two integration constants.

With these equations it is very easy to find the evolution of
perturbations in any phase. In fact, using the general expressions
of $\delta p$ and $\xi$, we can recast the set of equations in the
following form
\begin{equation}
X(\eta,k)=X^{(0)}(\eta,k)+\hat A(\eta,k) \int \hat
B(\eta,\partial_\eta, k) X(\eta,k) d\eta, \label{Eqgen}
\end{equation}
where $X=(\Psi,\zeta)$ is the vector of variables,
$X^{(0)}=(c_1(k)\mathcal{H}/a^2,c_2(k))$ is the vector containing
the decoupled solutions and $\hat A(\eta,k)$ and $\hat
B(\eta,\partial_\eta,k)$ are two coupling matrices that can be
deduced from Eqs. (\ref{Psiint}) and (\ref{zetaint}), once we use
the proper form of the sources as functions of $\zeta$ and $\Psi$.
The general solution to Eq. (\ref{Eqgen}) is
\begin{equation}
X=\sum\limits_{i=0}^\infty X^{(i)}, \label{Recsolsum}
\end{equation}
where $X^{(i)}$ is determined by $X^{(i-1)}$ as
\begin{equation}
X^{(i)}= \hat A(\eta,k) \int \hat B(\eta,\partial_\eta,k)
X^{(i-1)}(\eta,k) d\eta. \label{Recsol}
\end{equation}

So, starting from the decoupled solution $X^{(0)}$, we can construct the full
solution of the system. Of course, this only works for modes outside the horizon,
where we can stop to the first few terms of the series because the higher order
terms contain extra-factors of $k^2 \ll 1$ or $k^2 \eta^2 \ll 1$. Let us examine the
solution in each phase.

\subsection{Pre-bounce}

In the pre-bounce, the soulution assumes a very simple form. In
fact, there we have $\delta p=-2c_-^2 k^2\Psi/a^2$ and $\xi=0$,
while the explicit expression for all background functions in
terms of powers of $\eta$ is given by Eqs.
(\ref{PowLawa})-(\ref{PowLawq}). So, it is very easy to perform
all integrations. Discarding all numerical factors of order one,
we start from the decoupled solution, which assumes the following
form in the power law regime
\begin{eqnarray}
&& \Psi^{(0)}=k^{-2}c_1(k)|\eta|^{-1-2q_-} \\ &&
\zeta^{(0)}=c_2(k);
\end{eqnarray}
the first recursion gives
\begin{eqnarray}
&& \Psi^{(1)}=c_2(k) \\ && \zeta^{(1)}=c_1(k)|\eta|^{1-2q_-},
\end{eqnarray}
and the second recursion gives
\begin{eqnarray}
&& \Psi^{(2)}=c_1(k)|\eta|^{1-2q_-} \\ &&
\zeta^{(2)}=k^2c_2(k)|\eta|^2.
\end{eqnarray}
We can see that the second recursion gives the same modes already
present in the decoupled solution, multiplied by $k^2\eta^2$.
Since we are considering the evolution outside the horizon, this
factor is much smaller than one and we can safely neglect the
second and all higher order recursions. We are thus left with
\begin{eqnarray}
&& \Psi_-\sim k^{-2}c_1(k)|\eta|^{-1-2q_-} + c_2(k)\\ && \zeta_-
\sim c_1(k)\eta^{1-2q_-}+ c_2(k).
\end{eqnarray}
Comparing this solution with the expansion of the Hankel functions
for small arguments, given by Eqs. (\ref{zetapre}) and
(\ref{Psipre}), we can make the identifications $c_1(k)=k^\nu$,
$c_2(k)=k^{-\nu}$, recovering the two integration constants in
terms of the vacuum normalization.

\subsection{Bounce} \label{Sec Bounce}

Now let us turn to the core of our work. It must be said that the
specific shape of the solution during the bounce is not our
concern, since we just want to know the final spectrum of scalar
perturbations when they reenter the horizon. However, as we shall
see in the next subsection, in order to construct the post-bounce
solution, we need to evaluate the integrals appearing in Eq.
(\ref{Recsol}) in a domain covering the whole bounce phase, which
ranges from $\eta \simeq -1$ to $\eta \simeq 1$ in our normalized
units. It turns out that the evaluation of the bounce integrals is
very easy, since the $k$ and $\eta$ dependences are always
factored. In fact, the generic integral has the form
\begin{equation}
\int\limits_{-1}^1 f(k,\eta) d\eta= f_k(k)\int\limits_{-1}^1 f_\eta(\eta)d\eta = f_k(k). \label{bounceint}
\end{equation}

The last equality is crucial. It comes from the fact that the only
scales in our problem are $\eta$, $k$ and the bounce scale
$\eta_B$ that we have set to 1. Since we have factored the $k$
dependence, the only scale left in $f_\eta$ is $\eta$ itself. But
integrating over $\eta$ from $-1$ to $1$, whatever the dependence
of the primitive function on $\eta$, the final result will lose
the only scale it contained, and would thus be of order 1. It
cannot depend on any other scale, unless the bounce is actually
governed by two or more scales. In that case, we cannot solve the
integrals in such a trivial way, except in some special limits.

Since we are going to use several integrals over the bounce phase
in the next section, we give their explicit derivation in Appendix
A.

\subsection{Post-bounce} \label{Sec post-bounce}

In order to build the solution after the bounce, at every
recursion we must evaluate integrals of the form
\begin{equation}
\int\limits_{-\infty}^\eta f(k,\eta)d\eta.
\end{equation}
The integration domain can be split in three parts, covering the
pre-bounce, the bounce and the post-bounce phase, respectively. We
thus have
\begin{equation}
\int\limits_{-\infty}^\eta f(k,\eta)d\eta=
\int\limits_{-\infty}^{-1} f(k,\eta)d\eta  + \int\limits_{-1}^1
f(k,\eta)d\eta + \int\limits_{1}^\eta f(k,\eta)d\eta.
\end{equation}

For the pre-bounce integrals, it is sufficient to use the
pre-bounce solutions evaluated at $\eta \simeq -1$. For the bounce
integrals, we will exploit the explicit forms calculated in
Appendix A according to the prescription given in the previous
subsection. The post-bounce integrals can be easily evaluated
using the asymptotic background solution in the post-bounce phase,
where all background functions are replaced by powers of $\eta$.
For a generic function $f(k,\eta)=f_k(k)|\eta|^s$, we have
\begin{equation}
\int\limits_{1}^\eta f_k(k)|\eta|^s d\eta \simeq f_k(k)
\left[|\eta|^{s+1} + 1 \right],
\end{equation}
where we have discarded all numerical factors of order 1, as
usual.

We are now ready to construct the post-bounce solution. Once more,
we start from the decoupled solution, that in the post-bounce era
again assumes the explicit power-law form
\begin{eqnarray}
&& \Psi^{(0)}=k^{\nu-2}|\eta|^{-1-2q_+} \\ &&
\zeta^{(0)}=k^{-\nu}.
\end{eqnarray}
Then, the first recursion gives
\begin{eqnarray}
&& \Psi^{(1)}=|\eta|^{-1-2q_+}\left[ k^{-\nu}(|\eta|^{1+2q_+}+1) +
\int\limits_{-1}^1 f_\eta(\eta)\zeta^{(0)}+ \int\limits_{-1}^1
f_\eta(\eta)\xi^{(0)}d\eta
\right] \\%
&& \zeta^{(1)}=k^{\nu}(|\eta|^{1-2q_+}+1)+ \int\limits_{-1}^1
f_\eta(\eta)\delta p^{(0)} d\eta  + k^2\int\limits_{-1}^1
f_\eta(\eta)\xi^{(0)} d\eta,
\end{eqnarray}
where the bounce integrals can be read from the appendix.

The second recursion gives
\begin{eqnarray}
&& \Psi^{(2)}=|\eta|^{-1-2q_+}\left[
k^{\nu}(|\eta|^{1+2q_+}+|\eta|^2+1)\right. \nonumber \\ && \left.
+\left(\int\limits_{-1}^1 f_\eta(\eta)\delta p^{(0)} d\eta +
k^2\int\limits_{-1}^1 f_\eta(\eta)\xi^{(0)} d\eta
\right)(|\eta|^{1+2q_+}+1) \right. \nonumber \\ && \left. +
\int\limits_{-1}^1 f_\eta(\eta)\zeta^{(1)} d\eta  +
\int\limits_{-1}^1 f_\eta(\eta)\xi^{(1)}
d\eta \right] \\%
&& \zeta^{(2)}=k^{-\nu+2}(|\eta|^{1-2q_+}+|\eta|^{2}+1)\nonumber
\\ && + k^2 \left( \int\limits_{-1}^1
f_\eta(\eta)\zeta^{(0)}+ \int\limits_{-1}^1
f_\eta(\eta)\xi^{(0)}d\eta \right)(|\eta|^{1-2q_+}+1) \nonumber
\\ && +\int\limits_{-1}^1
f_\eta(\eta)\delta p^{(1)} d\eta + k^2\int\limits_{-1}^1
f_\eta(\eta)\xi^{(1)} d\eta.
\end{eqnarray}

No further recursions are needed, since all new terms are
negligible for superhorizon perturbations. At the end, according
to Eq. (\ref{Recsolsum}) we can sum up all contributions to build
the full post-bounce solution. A careful analysis reveals that
almost all terms coming from the bounce integrals are of the same
order of other terms already present or are negligible for
perturbations outside the horizon at the bounce. Then, the final
post-bounce solution takes the amazingly simple form
\begin{eqnarray}
& \Psi_+ & \sim k^{\nu-2}|\eta|^{-1-2q_+} + k^{-\nu} + k^{\nu} +G_0 k^{\nu-2} \label{Psipost}\\%
& \zeta_+ &\sim k^{\nu}\eta^{1-2q_+} + k^{-\nu} + k^{\nu}+ G_0
k^{\nu-2}, \label{zetapost}
\end{eqnarray}
where the constant $G_0$ is set to 1 if the corresponding operator
defined in Eq. (\ref{Gendp}) is non-vanishing; otherwise we have
$G_0=0$. In other words, this term is present only if $\delta p$
is directly proportional to $\Psi$ or its time derivatives,
without $k^2$ factors, during the bounce.

\subsection{Discussion}

The whole history of perturbations in a bouncing cosmology is
contained in Eqs. (\ref{zetapre}) and (\ref{Psipre}) for the
pre-bounce phase, and (\ref{Psipost}), (\ref{zetapost}) for the
post-bounce phase.

Let us start our discussion from the pre-bounce. The Bardeen
potential in the pre-bounce is dominated by the growing mode
$k^{\nu-2}|\eta|^{-1-2q_-}$, endowed with a red spectrum becoming
scale-invariant in the limit of slow contraction $q\rightarrow 0$.
This mode comes from the decoupled solution of Eq. (\ref{EqPsi}),
going as $\mathcal{H}/a^2$ and is destined to decay in the
post-bounce phase, where it takes the form
$k^{\nu-2}|\eta|^{-1-2q_+}$ in Eq. (\ref{Psipost}). There is also
a subdominant constant mode $k^{-\nu}$, which plays no role in
this phase. $\zeta$ couples to $\Psi$ through a Laplacian and thus
has a mode going as $k^{\nu}|\eta|^{1-2q_+}$ in Eq.
(\ref{zetapre}), where the potentially scale-invariant spectrum is
multiplied by a $k^2$ factor. This mode dominates for a fast
contraction $q_->1/2$ and is subdominant w.r.t. the constant mode
$k^{-\nu}$, coming from the decoupled solution of Eq.
(\ref{Eqzeta}), in the case of slow contraction $q_-<1/2$.

Now let us examine the post-bounce phase. The role of the bounce
is to upset the scalar perturbations for a finite time and to
determine the initial conditions for the post-bounce phase.
Immediately after the bounce, in Eq. (\ref{Psipost}) we have the
decoupled mode $k^{\nu-2}|\eta|^{-1-2q_+}$, carrying the
potentially scale-invariant spectrum, two constants modes
$k^{\nu}$ and $k^{-\nu}$ that are always present with a blue
spectrum, and possibly one more constant mode with the $k^{\nu-2}$
spectrum. {\it  This additional mode is present only if the
comoving pressure perturbation is proportional to $\Psi$ or its
time derivatives without Laplacians}. Recalling that we are in the
limit $k\ll 1$, this constant mode will dominate undisturbed until
horizon re-entry and will determine the spectrum of scalar
perturbations seeding the inhomogeneities of the universe.

In the other case (i.e. if $\delta p$ is related to $\Psi$ only
through Laplacians), the Bardeen potential is initially dominated
by its decaying mode. But at horizon re-entry, this mode has
decayed by a factor $\eta^{-1-2q_+}_{re}$, where $\eta_{re}$ is
the time of horizon re-entry. Setting $\eta_{re}\simeq k^{-1}$, we
have different possibilities. After a slow contraction in the
pre-bounce phase, the larger constant mode is $k^{-\nu}$. Then,
the decaying mode will still dominate only if $q_+<q_-$, i.e. if
the expansion in the post-bounce is slower than the pre-bounce
contraction. Otherwise, the constant mode dominates. However, in a
realistic cosmology the relevant modes for CMB enter the horizon
in the radiation or dust-dominated era, with $q_+=1$ or 2,
respectively. Then, the spectrum would be totally determined by
the blue constant mode $k^{-\nu}$, that in the very slow
contraction limit $q_- \ll 1$ gives a spectral index $n_s=3$. The
final possibility is that the universe has undergone a fast
pre-bounce contraction $q_->1/2$, so that the largest constant
mode is $k^{\nu}$. In this case, the decaying mode dominates when
the expansion is slow ($q_+<1/2$), while is negligible otherwise.

The modes dominating at horizon re-entry in all cases are
summarized by the following diagram

\begin{equation}
\left\{
\begin{array}{l}
G_0=1 \Longrightarrow \Psi\sim k^{\nu-2}\\%
G_0=0 \left\{
\begin{array}{l}
q_-<1/2 \left\{
\begin{array}{l}
q_+<q_- \Rightarrow \Psi\sim k^{\nu-2}\eta^{-1-2q_+} \\%
q_+>q_- \Rightarrow \Psi\sim k^{-\nu}
\end{array}
\right.
\\%
q_->1/2  \left\{
\begin{array}{l}
q_+<1/2 \Rightarrow \Psi\sim k^{\nu-2}\eta^{-1-2q_+} \\%
q_+>1/2 \Rightarrow \Psi\sim k^{\nu}
\end{array}
\right.
\end{array}
\right.
\end{array}
\right. \label{Diagram}
\end{equation}

At the same time, we see that there is a direct correspondence
between the modes present in $\Psi$ and those in $\zeta$, so that
an analogous discussion is valid also for $\zeta$. It is
interesting to note that a dependence on $\Psi$ or its time
derivatives in the anisotropic stress does not produce any
sizeable effects on the scalar perturbations. The anisotropic
stress enters without Laplacians in the equation for $\Psi$,
where, at most, it may renormalize the amplitude of the decaying
mode. In the equation for $\zeta$ it comes through a Laplacian,
thus being completely ineffective.

A very important {\it caveat} to keep in mind is that the presence
of a dependence on $\Psi$ in $\delta p$ is just a necessary
condition for the persistence of the $k^{\nu-2}$ spectrum in a
post-bounce constant mode. In fact, bounces with particular
symmetries may bring to special cancellations in the integrals,
erasing the lowest order contributions to the constant mode and
leaving space to higher order (most blue) contributions.

Another possibility is that the internal variables of the new
physics at the bounce may provide new modes characterized by
alternative spectra with a different physical origin w.r.t.
quantum fluctuations in the initial vacuum state. These modes may
contribute to the constant mode at the end of the bounce,
providing another possible alternative source for
phenomenologically acceptable scalar perturbations. However, in
such case, the phenomenology would completely rely on the high
energy physics at the bounce, whereas it would be preferable that
the final spectrum is as independent as possible of the unknown
bounce physics.

Our conclusions seem quite similar to those reached by Durrer and
Vernizzi in Ref. \cite{DurVer}, where the bounce is replaced by a
spacelike hypersurface and the evolution of perturbations through
the bounce is determined by Israel conditions. In fact, they find
that the persistence of the growing mode of the Bardeen potential
in the post-bounce era is very general, while its decay occurs
only if the bounce hypersurface coincides with the uniform energy
density hypersurface, which corresponds to force the surface
tension to be proportional to $k^2\Psi$. Reinterpreting the
surface tension as an integral of the pressure perturbation during
the bounce, this would correspond to tune $G_0=0$ in our
expression for $\delta p$. Thus, the two approaches seem in
complete agreement.

\section{Numeric testing on toy models}

Our golden rule for scalar perturbations through bouncing
cosmologies is based on a fully analytical derivation. However, it
is useful to visualize the main conclusions on some toy models of
bouncing cosmologies, so that they can be touched ``by hand''. In
order to make things as simple as possible, we have considered a
bounce with two perfect fluids: the first dominating the
pre-bounce and post-bounce eras and the second, being
characterized by negative energy density, representing the new
physics effective tensor inducing the bounce. For a bounce with
two perfect fluids, the Friedman equations are also integrable in
terms of hypergeometric functions \cite{Fin}, though we will not
use the analytical solutions. So, in this section, $\rho_a$ is the
energy density of the first fluid and $\rho_b$ is the energy
density of the second fluid. The two fluids have equations of
state $p_a=w_a \rho_a$, $p_b=w_b \rho_b$, with $w_a<w_b$, in order
to let the secondary fluid become negligible far from the bounce.
According to Eq. (\ref{PowLawq}), we have $q_-=q_+=q$. A slow
contraction followed by a slow expansion ($q<1/2$) takes place for
$w_a>1$, while a fast contraction followed by a fast expansion
($q>1/2$) takes place when $w_a<1$. As for all other models
already present in the literature, in this model the post-bounce
phase is the time-reversal of the pre-bounce phase. We are thus
unable to directly investigate a fast contraction followed by a
slow expansion or a slow contraction followed by a fast expansion.
Such asymmetric bouncing cosmologies require the implementation of
decay of the pre-bounce matter and creation of new matter at the
bounce, which would make the model much more complicated, though
not impossible to treat.

After the background is determined, we have to specify the form of
the energy-momentum tensor for perturbations. We choose
\begin{eqnarray}
&\delta p &= -\frac{2}{a^2}w_a k^2 \Psi
+3(w_b-w_a)(\rho_b+p_b)\zeta +G_0
(\rho_b+p_b) \Psi \\%
&\xi & =0.
\end{eqnarray}

In the comoving pressure perturbation $\delta p$ we have included
three terms. The first one is the classical adiabatic perturbation
of the first fluid, which is necessary to reproduce the ordinary
physics in the pre-bounce and post-bounce phases. The other two
terms are multiplied by the density of the secondary fluid, which
becomes negligible far from the bounce. The term proportional to
$\zeta$ is the same that one would have by the requirement that
the perturbations in the secondary fluid vanish in the gauge
$\psi=0$. This choice is not essential and has been made just for
computational reasons. The last term is a ``new physics'' term
directly proportional to $\Psi$. We will switch it on or off by
setting $G_0=1$ or $G_0=0$, respectively. Here we only present a
model with vanishing anisotropic stress, as this plays no role in
the final result, apart from a renormalization of some modes. Of
course, we have checked that this is actually the case, testing
some models (not to be presented here) with non-vanishing
anisotropic stress at the bounce.

With this choice for the energy-momentum tensor of perturbations,
the equations of motion are simply
\begin{eqnarray}
& \tilde{\zeta}'&+(1+3w_a)\mathcal{H}\tilde{\zeta}=-w_a
\mathcal{H} k^2 \Psi + \frac{G_0}{2} a^2\mathcal{H} (\rho_b+p_b)
\Psi
\\ & \Psi'&+\frac{2\mathcal{H}^2-\mathcal{H}'}{\mathcal{H}}\Psi
=\frac{\tilde{\zeta}}{\mathcal{H}}.
\end{eqnarray}
As anticipated before, it is safer to use
$\tilde{\zeta}=(\mathcal{H}^2-\mathcal{H}') \zeta$ in all
numerical applications, since this variable is guaranteed to be
regular throughout the bounce if any regular gauge exists.

\begin{figure}
\resizebox{10cm}{!}{\includegraphics{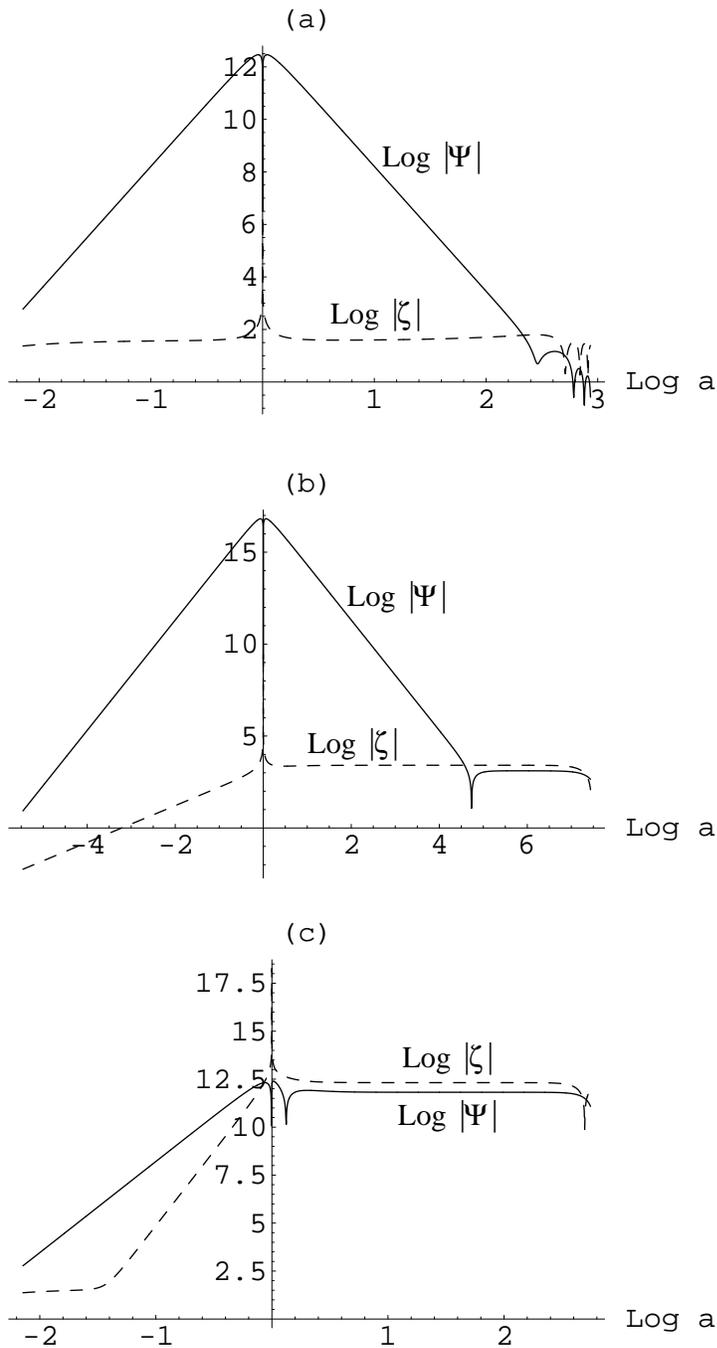}} \caption{
 The three possible behaviors of scalar perturbations in bouncing cosmologies.
 (a) $\delta p \sim k^2 \Psi$, with $q<1/2$ (slow
 contraction and slow re-expansion); (b) $\delta p \sim k^2 \Psi$, with $q>1/2$
 (fast contraction and fast re-expansion); (c) $\delta p \sim \Psi$, with any value of $q$.
 }
 \label{Fig Num}
\end{figure}

Imposing the vacuum normalization as the initial condition, our
toy model is well defined and we can choose different values of
$w_a$, $w_b$ and $G_0$ in order to test the conclusions reached
analytically. The results are summarized in Fig. \ref{Fig Num},
where we have plotted $\log |\Psi|$ and $\log |\zeta|$ as
functions of $\mathrm{Sign}[\eta]\log a$, which is now a common
practice for the visualization of the behavior of perturbations in
bouncing cosmologies \cite{AllWan,BozVen}.

For Fig. \ref{Fig Num}a, we have chosen $w_a=1.5$ (implying
$q=0.36$), $w_b=2.5$ and $G_0=0$. For Fig. \ref{Fig Num}b, we have
$w_a=1/3$ (implying $q=1$), $w_b=1$ and $G_0=0$. For Fig. \ref{Fig
Num}c, we have $w_a=1.5$, $w_b=2.5$ and $G_0=1$. These three cases
illustrate almost everything we can expect in a bouncing
cosmology. The spectral index $n_s$ of the perturbations can be
evaluated at any time, evolving at least two Fourier modes with
different $k$ and fitting the power law $k^{n_s-1}$. The exponent
of the $\eta$-dependence is extracted by the logarithmic time
derivative of the modulus of the perturbation. In this way we can
compare the analytical predictions for the $k$ and $\eta$
dependences with the numerical results in all phases.

\begin{figure}
\resizebox{10cm}{!}{\includegraphics{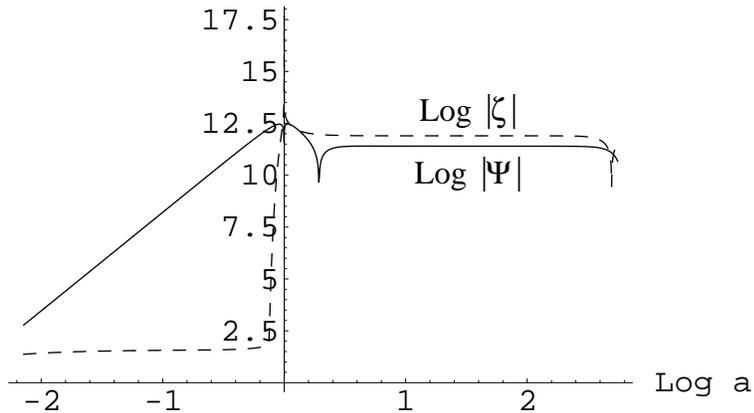}} \caption{
 The same case as Fig. \ref{Fig Num}c with a different relation between $\delta p$ and $\Psi$.
 Here we have chosen $\delta p \sim (\rho_b+p_b) e^{1/\rho_b}\Psi$.
 }
 \label{Fig Exp}
\end{figure}

For the pre-bounce, we just have to distinguish $q<1/2$ (slow
contraction) from $q>1/2$ (fast contraction). In both cases,
$\Psi$ is dominated by the growing mode with the red spectrum
$k^{\nu-2}|\eta|^{-1-2q_-}$, as can be seen in all three plots in
Fig. \ref{Fig Num}. $\zeta$ is dominated by the constant mode
$k^{-\nu}$ in a slow contraction (Fig. \ref{Fig Num}a) and by the
growing mode $k^{\nu}|\eta|^{1-2q_-}$ in a fast contraction (Fig.
\ref{Fig Num}b). However, when $G_0=1$, we see in Fig. \ref{Fig
Num}c that the pre-bounce constant mode (we are in a slow
contraction) is taken over by a mode going as
$k^{\nu-2}|\eta|^{-2-3(1+w_b)q_-}$ at some time before the bounce.
This mode is induced by the direct coupling of $\zeta$ to $\Psi$
that starts to be effective during the approach to the bounce. In
fact, in the model presented here, even if the term proportional
to $\Psi$ in $\delta p$ is suppressed by $(\rho_b+p_b)$, $\Psi$
grows so large that the $G_0$-mode starts to dominate already some
time before the bounce. Indeed, choosing a function going to zero
faster than $(\rho_b+p_b)$ far from the bounce, we can delay the
growth of $\zeta$ as much as we like. For example, choosing
$(\rho_b+p_b) e^{1/\rho_b}$, the growth of $\zeta$ practically
starts right at the bounce, but the post-bounce result is
essentially unchanged, as shown in Fig. \ref{Fig Exp}.

The post-bounce scalar perturbations essentially depend on the
presence of a direct proportionality between $\delta p$ and
$\Psi$. If we set $G_0=1$, we see from Fig. \ref{Fig Num}c that
immediately after the bounce both $\zeta$ and $\Psi$ are dominated
by a constant mode carrying the pre-bounce spectrum of $\Psi$,
i.e. $k^{\nu-2}$. We want to stress that this outcome is
completely independent of the expansion and contraction rates and
of the specific form of the background function that appears in
the coupling of $\delta p$ to $\Psi$. Choosing $(\rho_b+p_b)$,
$(\rho_b+p_b) e^{1/\rho_b}$ or any other function gives exactly
the same result, apart from numerical factors of order 1.

If the $G_0$-mode is completely absent, then the pre-bounce
growing mode of the Bardeen potential is entirely converted into a
decaying mode, leaving no trace in the constant modes. Then, at
horizon re-entry it may still dominate or it may be negligible
w.r.t. some constant mode, depending on the contraction and
expansion rates. In Fig. \ref{Fig Num}a we have a slow contraction
followed by an expansion with the same rate. We see that at
horizon re-entry the decaying mode of $\Psi$ is of the same order
as the constant mode of $\zeta$, in agreement with the diagram
(\ref{Diagram}), where this case is marginal between the dominance
of $k^{-\nu}$ and the dominance of the decaying mode. In Fig.
\ref{Fig Num}b we have a fast contraction followed by an expansion
with the same rate. In agreement with the diagram (\ref{Diagram}),
the decaying mode is taken over by a constant mode with a spectrum
$k^{\nu}$.

\section{Remarks on bounces induced by spatial curvature and extra-dimensions}

In hypothesis (ii) we have excluded the presence of spatial
curvature. Indeed, this would introduce another scale in our
problem and in order to be compatible with hypothesis (iii), we
have two possibilities.

The first is that the curvature scale is very low compared to the
bounce scale. Then the universe is practically spatially flat and
we can neglect the spatial curvature to follow the treatment of
this paper.

The second is that the curvature itself determines the bounce so
that the curvature scale at the bounce is actually what we have
called $\eta_B=1$. Spatial curvature may induce a cosmological
bounce only if it is positive, because it enters the Friedman
equation as a perfect fluid with negative energy density and
equation of state $w=-1/3$. The remaining source-matter may keep
satisfying $\rho+p>0$, since the effective NEC violation is
performed by the spatial curvature. In order to let the curvature
become negligible before and after the bounce, we need $\ddot
a>0$, i.e. decelerated contraction and accelerated expansion. For
example, the pre-bounce and the post-bounce phases may be
dominated by sources with equations of state $w<-1/3$. This
contradicts our hypothesis (iv) and the initial conditions would
be problematic without another early transition from an
accelerated contraction in the far pre-bounce. This possibility
was considered by several authors \cite{K>0} as a toy model for a
bouncing cosmology.

However, even relaxing hypothesis (iv), there is a structural
difference between a bounce induced by spatial curvature and all
other bounces with negligible spatial curvature scale. In fact,
the Laplace-Beltrami operator in a positively curved space has a
discrete spectrum $k^2=n(n+2)$ (recall that the curvature scale at
the bounce has been set to 1). Then, all modes are in the
situation $k>1$, that means that all modes are {\it inside} the
horizon at the bounce. Then our recursive solution of the integral
equation would require an infinite number of terms and is
completely useless. In practice, all modes re-enter the horizon
during the pre-bounce deflation and oscillate in the approach to
the bounce. During the inflationary phase after the bounce, they
exit the horizon again, stopping the oscillations. This behavior
has been put in evidence by Deruelle \cite{Deruelle}, who has
described the oscillations analytically by hypergeometric
functions. Then, it is not surprising that mode mixing after the
oscillations is a generic outcome of such scenarios.

We have to conclude that bounces with positive curvature have a
substantially different physics and cannot be used as a step to
investigate bounces in spatially flat universes. Then it is
difficult to believe that they could be useful for any physically
motivated models.

However, there is one remarkable feature of spatial curvature that
deserves to be mentioned. We have already noted that spatial
curvature behaves as a perfect fluid with equation of state
$p=-\rho/3$ at the background level. At the linear perturbation
level, we may try to recast all equations in such a way that they
look like Eqs. (\ref{Eqdrho})-(\ref{EqPsi}) with some effective
energy-momentum tensor containing the curvature terms. After some
manipulation, we find the following form for the perturbations
induced by the spatial curvature in the comoving gauge
\begin{eqnarray}
&& \delta \rho= -\frac{6}{a^2}\Psi \\%
&& \delta p= \frac{2}{a^2}\Psi
+\frac{2}{a^2\mathcal{H}}(\Psi'-\zeta')
- \frac{1}{a^2\mathcal{H}}k^2(\Psi-\zeta)\\%
&& \xi=\frac{1}{a^2\mathcal{H}^2}(\zeta-\Psi).
\end{eqnarray}

Spatial curvature thus represents a very interesting example of
some ``new physics'' with a pressure perturbation directly
proportional to $\Psi$ and its time derivative without the
Laplacian. So, even if spatial curvature is not useful for
realistic bouncing cosmologies by itself, it encourages the
conjecture that the new physics driving the bounce may contain
terms directly proportional to the Bardeen potential in the
effective energy-momentum tensor of perturbations.

Perhaps an even more interesting surprise comes out from
extra-dimensions. In some particular cases, in fact, it is
possible to recast D-dimensional field equations in the form of
4-dimensional Friedman equations with some effective
energy-momentum tensor containing the dynamics of the
extra-dimensions. One example is the case in which the spacetime
is factored into a 4-dimensional FRW metric and an internal
homogeneous manifold, where none of the metric components depends
on the coordinates spanning the internal space. For example,
consider a five-dimensional spacetime with the metric
\begin{eqnarray}
&ds^2=&a^2(\eta)\left\{ (1+2\phi) d\eta^2-2B_{,i} d\eta dx^i
-\left[(1-2\psi)\delta_{ij}+2E_{,ij}\right]dx^i dx^j \right\}
 \nonumber \\&&-b^2(\eta)(1+2\Gamma)dy^2-2Wdtdy,
\end{eqnarray}
with the assumption that all scalar perturbations do not depend on
the internal coordinate $y$ (i.e. we are only including zero modes
and neglecting excitations of the extra-dimensions). This metric
was used by Battefeld, Patil and Brandenberger \cite{Extra} to
study bounces with extra-dimensions. It is easy to show that the
background equations have the form of 4-dimensional Friedman
equations with an effective contribution to the total
energy-momentum tensor coming from the extra-dimension
\begin{equation}
\rho_{extra}=-3\mathcal{H} a^{-2}\frac{b'}{b}, ~~~~~~
p_{extra}=a^{-2}\left( \frac{b''}{b}+ \mathcal{H} \frac{b'}{b}
\right),
\end{equation}
where the (yy) component of the Einstein equations has been used
to eliminate ${T^5}_5$.

In the same way, examining the perturbations equations, we can
re-obtain Eqs. (\ref{Eqdrho})-(\ref{EqPsi}) with effective
$\delta\rho_{extra}$, $\delta p_{extra}$ and $\xi_{extra}$
containing terms depending on $\Psi$, $\zeta$ $\Gamma$ and their
derivatives ($W$ decouples from other perturbations). Thus,
$\Gamma$ plays the role of internal variable of the model. Its
dynamical equation is given by the perturbed (yy) component, with
$\delta {T^5}_5$ as a source.

We will not present the complete expressions of the perturbed
effective energy-momentum tensor here, but the important lesson is
that the presence of terms directly proportional to $\Psi$ in
$\delta p$ is generic in this class of models as well. Then,
bounces with extra-dimensions may be able to transfer the
pre-bounce spectrum of $\Psi$ to the post-bounce. This is in
agreement with the conclusions of Ref. \cite{Extra}.

The fact that extra-dimensions perturbations contribute to the
pressure perturbation was already noted by Giovannini \cite{Gio1}.
Coherently with the behavior of scalar perturbations, he also
finds that vector perturbations in multi-dimensional bouncing
universes preserve the red spectrum acquired in the contraction
phase, in contrast with 4-dimensional bouncing universes, where
the red spectrum is only transferred to the decaying mode
\cite{Gio2}.

Nevertheless, a complete analysis of scalar perturbations in
multi-dimensional bouncing universes that also takes into account
non-zero modes is still missing. Their relevance certainly
deserves some investigation, but they cannot fit into the
effective four-dimensional picture presented in this paper.
Further extensions of our methodology are certainly possible.

\section{Conclusions}

In this paper, we have presented the general solution for scalar
perturbations in bouncing cosmologies, i.e. models of the universe
where the present expansion is preceded by a contraction phase.
This solution has been derived solving integral equations where
the only assumption on the bounce is that it depends on a single
physical scale. The ``new physics'' ruling the bounce has been
encoded in an effective energy-momentum tensor, for which we have
made no {\it a priori} assumption.

The final spectrum of the Bardeen potential in the expansion phase
depends on the presence or absence in the new physics
energy-momentum tensor of a direct relation between the comoving
pressure perturbation and the Bardeen potential, not mediated by a
Laplacian, as in all known forms of ordinary matter. If such term
is present, the pre-bounce growing mode of the Bardeen potential
persists in the post-bounce constant mode and it is possible to
have a phenomenologically acceptable scale-invariant spectrum in
Ekpyrotic-like cosmologies (i.e. models with very slow contraction
in the pre-bounce). If such term is absent, the pre-bounce growing
mode of the Bardeen potential decays in the post-bounce and
becomes negligible w.r.t. a blue constant mode, unless the
post-bounce expansion is slower than the pre-bounce contraction.

The negative conclusion reached by all regular models of
4-dimensional spatially flat bounces studied up to now is due to
the absence of such dependence in the energy-momentum tensor.
Then, in order to support our conclusion, we have presented a
numerical study of a toy model with $\delta p \propto \Psi$ which
transfers the spectrum of the pre-bounce growing mode of $\Psi$ to
the post-bounce growing mode.

Bounces induced by spatial curvature are structurally different
from spatially flat bounces, since all modes are inside the
horizon for some time interval centered at the bounce. Our
treatment cannot be applied to this distinct category of bounces,
whose study is anyway useless for the understanding of bounces in
spatially flat cosmologies. However, spatial curvature provides a
remarkable example of an effective energy-momentum tensor where
the pressure perturbation is proportional to the Bardeen potential
without Laplacians. The same key property is also possessed by
models of bounces with extra-dimensions where the dependence on
the coordinates spanning the internal space is neglected. These
multi-dimensional models can fit into our effective
four-dimensional scheme. However, before drawing a positive
conclusion, a more rigorous and complete treatment (taking into
account non-zero modes) should be envisaged for multi-dimensional
bounces.

It thus seems not impossible to conjecture a form for the new
physics driving the bounce that realizes the Ekpyrotic dream. The
problem now is not in matching conditions or any technical issues
about the evolution of perturbations, which are now definitively
settled by this work. The missing ring of the chain is just the
existence of some explicit model with the required property of the
effective energy-momentum tensor and that is physically motivated
by a consistent high energy theory like string or M-theory. When
this explicit model is found, then it will be possible to say that
a real alternative to standard inflation exists.

On the other hand, a more prudent attitude suggests the
possibility that there might exist some important physical reasons
precluding the possibility that an effective energy-momentum
tensor of the required form is allowed at the bounce and then the
way for bouncing cosmologies would be definitively barred.
However, it seems very hard to us to find any physical reason to
forbid an effective pressure perturbation directly proportional to
the Bardeen potential. Moreover, the explicit examples of spatial
curvature and extra-dimensions seem to indicate that this
possibility is actually open for high energy theories.

\ack

I am very grateful to Gabriele Veneziano for encouraging me to
complete this work and for illuminating discussions. I also thank
the CERN Theory Unit for hospitality. This work was partially
funded by Centro studi e ricerche ``Enrico Fermi'', Rome, Italy. I
also acknowledge useful discussions with Shini Tsujikawa, Ghazal
Geshnizjani and Thirtabir Biswas.

\appendix

\section{Bounce integrals}

The calculation of all bounce integrals according to the
prescription of Sect. \ref{Sec Bounce} is straightforward, though
a bit boring. Though we need to calculate a relatively large
number of terms for the second recursion, luckily most of them are
similar to other terms coming from the pre-bounce or post-bounce
phases, so that they are irrelevant for the final post-bounce
spectrum.

The starting point is the decoupled solution
\begin{eqnarray}
&& \Psi^{(0)}=k^{\nu-2}\frac{\mathcal{H}}{a^2} \\ &&
\zeta^{(0)}=k^{-\nu}.
\end{eqnarray}
Note that we do not specify the shape of $a$ and $\mathcal{H}$
during the bounce phase. Then, indicating by $f_\eta(\eta)$ a
generic function of $\eta$, the integrals involved in the first
recursion are
\begin{eqnarray}
&& \int\limits_{-1}^1 f_\eta(\eta)\zeta^{(0)} d\eta=k^{-\nu} \\ &&
\int\limits_{-1}^1 f_\eta(\eta)\xi^{(0)}
d\eta=\sum\limits_{n=0}^{\infty} \left( J_n k^{-\nu+2n}+K_n
k^{\nu-2+2n} \right)\\ && \int\limits_{-1}^1 f_\eta(\eta)\delta
p^{(0)} d\eta=\sum\limits_{n=0}^{\infty} \left( F_n
k^{-\nu+2n}+G_n k^{\nu-2+2n} \right),
\end{eqnarray}
where the first and the second integrals enter $\Psi^{(1)}$, while
$\zeta^{(1)}$ needs the second (multiplied by $k^2$) and the
third. We have used the expressions (\ref{Gendp}) and
(\ref{Genxi}) for $\delta p^{(0)}$ and $\xi^{(0)}$. By the symbols
$F_n$, $G_n$, $J_n$, $K_n$ we generically indicate the results of
all integrals involving the corresponding operators defined in
Eqs. (\ref{Gendp}) and (\ref{Genxi}). Of course, these integrals,
if present, are of order 1, according to the argument in Eq.
(\ref{bounceint}). We will keep the symbols $F_n$, $G_n$, $J_n$,
$K_n$ just to remember that these terms are present only if the
corresponding operators are not zero. At the end it will be
sufficient to put these symbols to 1 or 0 to consider models where
these operators are present or absent, respectively.

In the second recursion we need the integrals
\begin{eqnarray}
&& \int\limits_{-1}^1 f_\eta(\eta)\zeta^{(1)} d\eta=
  \sum\limits_{n=0}^{\infty} \left( F_n k^{-\nu+2n}+G_n k^{\nu-2+2n}  + J_n k^{-\nu+2n+2}+K_n k^{\nu+2n} \right) \\
&& \int\limits_{-1}^1 f_\eta(\eta)\xi^{(1)} d\eta=
\sum\limits_{n=0}^{\infty} K_n k^{-\nu+2n} +
\sum\limits_{n,m=0}^\infty \left( J_m F_n k^{-\nu+2m+2n} + J_m G_n
k^{\nu-2+2m+2n} \right. \nonumber \\ && \left. + J_m J_n
k^{-\nu+2+2m+2n} + J_m K_n k^{\nu+2m+2n} \right. \nonumber \\ &&
\left. + K_m J_n k^{-\nu+2m+2n}+ K_m K_n k^{\nu-2+2m+2n} \right)
\\ && \int\limits_{-1}^1 f_\eta(\eta)\delta p^{(1)} d\eta=
\sum\limits_{n=0}^{\infty} G_n j^{-\nu+2n}  +
\sum\limits_{n,m=0}^\infty \left( F_m F_n k^{-\nu+2m+2n}+F_m G_n
k^{\nu-2+2m+2n} \right. \nonumber \\ && \left. + F_m J_n
k^{-\nu+2+2m+2n}+ F_m K_n k^{\nu+2m+2n} \right. \nonumber \\ &&
\left. + G_m J_n k^{-\nu+2m+2n}+ G_m K_n k^{\nu-2+2m+2n} \right).
\end{eqnarray}

These integrals are those used in Sect. \ref{Sec post-bounce} to
derive the post-bounce solution for scalar perturbations.


\section*{References}

\begin{thebibliography}{}

\bibitem{PBB} Veneziano G, 1991 {\it Phys. Lett.} B {\bf 265}, 287
\\ Gasperini M and Veneziano G, 1993 {\it Astropart. Phys.} {\bf 1}, 317 \\
Gasperini M and Veneziano G, 2003 {\it Phys. Rep.} {\bf 373}, 1

\bibitem{OtherBounce} Constantinidis C P, Fabris J C, Furtado R G and
Picco M, 2000 {\it Phys. Rev.} D {\bf 61}, 043503 \\ Kogan I I,
Mouslopoulos S, Papazoglou A and Ross G G, 2001 {\it Phys. Rev.} D
{\bf 64}, 124014 \\ Mukherji S and Peloso M, 2002 {\it Phys.
Lett.} B {bf 547}, 297 \\ Kachru S and McAllister L, 2003 {\it
JHEP} {\bf 0303}, 018 \\ Shtanov Y and Sahni V, 2003 {\it Phys.
Lett.} B {\bf 557}, 1 \\ Foffa S, 2003 {\it Phys. Rev.} D {\bf
68}, 043511 \\ Veneziano G, 2004 {\it JCAP} {\bf 0403}, 004 \\
Setare M R, 2004 {\it Phys. Lett.} B {bf 602}, 1 \\ Date G and
Mortuza Hossain G, 2005 {\it Phys. Rev. Lett.} {\bf 94}, 011302 \\
Rinaldi M and Watts P, 2005 {\it JCAP} {\bf 0503}, 006; Herdeiro C
A R and Sampaio M, 2005 {\it Preprint} hep-th/0510052 \\ Carloni
S, Dunsby P K S and Solomons D, 2005 {\it Preprint} gr-qc/0510130

\bibitem{Ekp} Khouri J, Ovrut B A, Steinhardt P J and Turok N, 2002 {\it Phys. Rev.}
D {\bf 66}, 046005 \\ Gratton S, Khoury J, Steinhardt P J and
Turok N, 2004 {\it Phys. Rev.} D {\bf 69}, 103505 \\ Tolley A J,
Turok N and Steinhardt P J, 2004 {\it Phys. Rev.} D {\bf 69},
106005

\bibitem{FinBra} Finelli F and Brandenberger R, 2002 {\it Phys. Rev.} D {\bf 65}, 103522

\bibitem{Biswas} Biswas T, Mazumdar A and
Siegel W, 2005 {\it Preprint} hep-th/0508194

\bibitem{PBBPert} Brustein R, Gasperini M, Giovannini M, Mukhanov V and
Veneziano G, 1995 {\it Phys. Rev.} D {\bf 51}, 6744

\bibitem{BGGV} Enqvist K and Sloth M S, 2002 {\it Nucl. Phys.} B {\bf 626},
395\\ Lyth D H and Wands D, 2002 {\it Phys. Lett.} B {\bf 524}, 5
\\ Bozza V, Gasperini M, Giovannini M and Veneziano G, 2002 {\it
Phys. Lett.} B {\bf 543}, 14 \\ Bozza V, Gasperini M, Giovannini M
and Veneziano G, 2003 {\it Phys. Rev.} D {\bf 67}, 063514

\bibitem{Others}  Lyth D, 2002 {\it Phys. Lett.} B {\bf 524}, 1 \\
Brandenberger R and Finelli F, 2001 {\it JHEP} {\bf 0111}, 056 \\
Lyth D, 2002 {\it Phys. Lett.} B {\bf 526}, 173 \\ Hwang J, 2002
{\it Phys. Rev.} D {\bf 65}, 063514; Tsujikawa S, 2002 {\it Phys.
Lett.} B {\bf 526}, 179 \\ Martin J, Peter P, Pinto-Neto N and
Schwarz D J, 2002 {\it Phys. Rev.} D {\bf 65}, 123513 \\ Hwang J
and Noh H, 2002 {\it Phys. Lett.} B {\bf 545}, 207 \\ Martin J,
Peter P, Pinto-Neto N and Schwarz D J, 2003 {\it Phys. Rev.} D
{\bf 67}, 028301 \\ Tsujikawa S, Brandenberger R and Finelli F,
2002 {\it Phys. Rev.} D {\bf 66}, 083513 \\ Cartier C, Durrer R
and Copeland E, 2003 {\it Phys. Rev.} D {\bf 67}, 103517 \\ Peter
P, Pinto-Neto N and Gonzalez D A, 2003 {\it JCAP} {\bf 0312}, 003
\\ Creminelli P, Nicolis A and Zaldarriaga M, 2005 {\it Phys. Rev.}
D {\bf 71}, 063505 (2005)

\bibitem{DurVer} Durrer R and Vernizzi F, 2002 {\it Phys. Rev.} D {\bf 66}, 083503

\bibitem{Cartier} C. Cartier, 2004 {\it Preprint} hep-th/0401036

\bibitem{GGV} Gasperini M, Giovannini M and Veneziano G, 2003 {\it
Phys. Lett.} B {\bf 569}, 113 \\ Gasperini M, Giovannini M and
Veneziano G, 2004 {\it Nucl. Phys.} B {\bf 694}, 206

\bibitem{AllWan} Allen L and Wands D, 2004 {\it
Phys. Rev.} D {\bf 70}, 063515

\bibitem{BozVen} Bozza V and Veneziano G, 2005 {\it Phys.
Lett.} B {\bf 625}, 177 \\ Bozza V and Veneziano G, 2005 {\it
JCAP} {\bf 0509}, 007

\bibitem{PetPin} Peter P and Pinto-Neto N, 2002 {\it Phys. Rev.} D {\bf 66},
063509

\bibitem{Fin} Finelli F, 2003 {\it JCAP} {\bf 0310}, 011

\bibitem{BatGes} Battefeld T J and Geshnizjani G, 2005 {\it Preprint} hep-th/0503160

\bibitem{K>0} Hwang J and Noh H, 2002 {\it Phys. Rev.} D {\bf 65},
124010 \\ Gordon C and Turok N, 2003 {\it Phys. Rev.} D {\bf 67},
123508 \\ Martin J and Peter P, 2003 {\it Phys. Rev.} D {\bf 68},
103517 \\ Martin J and Peter P, 2004 {\it Phys. Rev. Lett.} {\bf
92}, 061301 \\ Martin J and Peter P, 2004 {\it Preprint}
gr-qc/0406062

\bibitem{Deruelle} Deruelle N, 2004 {\it Preprint} gr-qc/0404126
\\ Deruelle N and Streich A, 2004 {\it Phys. Rev. } D {\bf 70},
103504

\bibitem{Starobinsky} Tsujikawa S, {\it private communication,
reporting A. Starobinsky's opinion}

\bibitem{Extra} Battefeld T J, Patil S P and Brandenberger R H,
2005 {\it Preprint} hep-th/0509043

\bibitem{Gio1} Giovannini M, 2005, {\it Class. Quant. Grav.} {\bf 22}, 2201

\bibitem{Gio2} Giovannini M, 2004, {\it Phys. Rev.} D {\bf 70}, 103509

\end{thebibliography}
\end{document}